\documentclass[conference,final]{IEEEtran}
\IEEEoverridecommandlockouts

\usepackage[T1]{fontenc} 
\usepackage[utf8]{inputenc}
\usepackage{amsmath}
\usepackage[cmintegrals]{newtxmath}
\usepackage{bm} 

\usepackage{hyperref}
\hypersetup{
    colorlinks=true,
    linkcolor=blue,
    filecolor=magenta,      
    urlcolor=cyan,
}
\usepackage{cite}
\usepackage{amsmath,amsfonts}
\usepackage{algorithmic}
\usepackage{graphicx}
\usepackage{textcomp}
\usepackage{xcolor}
\usepackage{todonotes}
\usepackage{xspace}
\usepackage{tikz}
\usepackage{lipsum}

\newcommand\copyrighttext{%
  \footnotesize \textcopyright 2021 IEEE. Personal use of this material is permitted.
  Permission from IEEE must be obtained for all other uses, in any current or future 
  media, including reprinting/republishing this material for advertising or promotional 
  purposes, creating new collective works, for resale or redistribution to servers 
  or lists, or reuse of any copyrighted component of this work in other works. 
  DOI: \href{https://doi.org/10.1109/SEH52539.2021.00014}{10.1109/SEH52539.2021.00014}
}

\newcommand{\tooltxt}{SOD\xspace}
\newcommand{\tool}{{\sf \tooltxt}\xspace}

\newcommand{\circled}[1]{\textcircled{\small{#1}}}

\def\BibTeX{{\rm B\kern-.05em{\sc i\kern-.025em b}\kern-.08em
    T\kern-.1667em\lower.7ex\hbox{E}\kern-.125emX}}
\begin{document}

\title{Wandering and getting lost: the architecture of an app activating local
communities on dementia issues
}

\author{\IEEEauthorblockN{Nicklas Sindlev Andersen}
\IEEEauthorblockA{\textit{University of Southern Denmark}\\
Odense, Denmark \\
\emph{sindlev@imada.sdu.dk}}
\and
\IEEEauthorblockN{Marco Chiarandini} \IEEEauthorblockA{\textit{University of
Southern Denmark}\\
Odense, Denmark \\
marco@imada.sdu.dk}
\and
\IEEEauthorblockN{Jacopo Mauro} \IEEEauthorblockA{\textit{University of Southern
Denmark}\\
Odense, Denmark \\
mauro@imada.sdu.dk}}

\maketitle

\begin{tikzpicture}[remember picture,overlay]
    \node[anchor=south,yshift=10pt] at (current page.south) {\fbox{\parbox{\dimexpr\textwidth-\fboxsep-\fboxrule\relax}{\copyrighttext}}};
\end{tikzpicture}%

\begin{abstract}
    We describe the architecture of Sammen Om Demens (\tool), an
application for portable devices aiming at helping persons with
dementia when wandering and getting lost through the involvement of
caregivers, family members, and ordinary citizens who volunteer.

To enable the real-time detection of a person with dementia that has
lost orientation, we transfer location data at high frequency from a
frontend on the smartphone of a person with dementia to a backend
system. The backend system must be able to cope with the high
throughput data and carry out possibly heavy computations for the
detection of anomalous behavior via artificial intelligence
techniques. This sets certain performance and architectural
requirements on the design of the backend.

In the paper, we discuss our design and implementation choices for the
backend of \tool that involve microservices and serverless services to
achieve efficiency and scalability. We give evidence of the achieved
goals by deploying the \tool backend on a public cloud and measuring
the performance on simulated load tests.

\end{abstract}


\section{Introduction} \label{sec:introduction}
An increasing number of people in the world are estimated to be
affected by dementia \cite{war2015}. As such, there is an increasing
effort to conceive social and technological innovations to help people
affected by dementia syndrome along with their relatives, friends, and
communities~\cite{wen2019}. Wandering, disorientation, and getting lost
are common in people with dementia, even in the early stage, and there
is a considerable effort to provide technological solutions aimed at
alleviating the troubles related to these
situations~\cite{Hassan2019,Ray2020}.

In line with this effort, we developed an application for portable
devices, dubbed ``Sammen Om Demens'' (\tool), in the context of a
Danish municipality. The goal of
\tool is to create awareness among ordinary citizens and to involve them in
helping persons affected by dementia, ultimately alleviating the
anxiety of these latter and their closest caregivers. The main
functionality of \tool that is expected to achieve this goal is a
software component that will be able to automatically detect episodes
of wandering and getting lost.

In this paper, we focus on the architecture of \tool that relies on
a \emph{frontend} software deployed on portable devices
(i.e.,~smartphones) and a \emph{backend} system that runs in the
cloud. The frontend collects and sends location data at a high
frequency to the backend system that processes the information
received.  The backend is designed to allow the integration of the
software component that automatically detects wandering and getting
lost situations. For this reason, the backend is required to be able
to process an expected high throughput of data efficiently and
reliably, while being structurally flexible to allow to easily deploy
artificial intelligence techniques for the detection of persons with
dementia that get lost. It is thus vital to design the backend of the
system that can operate efficiently and reliably, easily scaling as the 
number of users of \tool increase. Our
solution has been to adopt microservices and serverless
services. Here, we will argue and give evidence that these types of architectural
components offer several advantages, among them indeed scalability and
architectural flexibility~\cite{microservices,serverless}.
%
%
In the paper, we restrict ourselves to describe and evaluate the
microservices and serverless architecture and its performance. Testing the app 
with real users, comparing it against other approaches (see
e.g.~\cite{9274637,BERROCAL2017106}), or detailing
the automatic detection of anomalous behavior is left as future work for a 
follow-up article.

The paper is structured as follows. In Section~\ref{sec:sammen_om_demens}, we
give more details on the features of \tool, that need to be supported by the
backend system. In Section~\ref{sec:preliminaries}, we describe the underlying
paradigms and technologies that we used in the development of the backend
system. In Section~\ref{sec:implementation}, we describe the implemented
infrastructure before presenting its validation in
Section~\ref{sec:performance_evaluation}. We conclude in
Section~\ref{sec:related_literature_conclusions} discussing related work and
future research plans.

\section{Sammen Om Demens (SOD)} \label{sec:sammen_om_demens}
In this section, we give an overview of the features of \tool that
need to be supported by the backend system.

The backend system accommodates three user types: i)~users with dementia (from
light to medium cases) who can use to a certain degree a smartphone and remember
to bring it with them ii) closest caregivers and relatives linked to one or more
users with dementia and iii) ordinary citizens, not linked to any specific user,
who volunteer to help. Relatives to a person with dementia are handled as
volunteers but have enhanced functionalities towards the user with dementia to
whom they are linked.
For the final users, the app consists of three components each providing a
different functionality: a knowledge base providing information about the local
organization of the municipality around dementia, a help component providing
functionalities to activate relatives and near volunteers in cases of need, and
a recreational activity calendar with the additional feature of proposing
accompanying persons.

The \textit{help component} is concerned with the detection of persons with
dementia that exhibit some form of anomalous behavior such as becoming lost or
being confused about their location. Through the app, it is possible for a user
with dementia to directly trigger an alarm if he/she became lost and wishes
assistance. Alternatively, a relative of a user with dementia can detect the
anomaly through the enhanced capability of this user type to track the location
of their linked users on a map. They can then trigger the alarm to contact their
linked user and activate volunteers. Most interestingly, the same alarm can also
be triggered \emph{automatically} by the app if the backend system detects
anomalous behavior in the stream of location data sent by the frontend.
Eventually, if an alarm is directly or indirectly triggered, the backend system
has the task of sending out notifications to the nearest available volunteers
and to the available relative of the user with dementia. In case of more
relatives, the one registered with the highest priority is chosen. When the
notification is received by a user, it triggers an alarm sound to warn the user
to react fast. The volunteer users that receive the notification have the
choice to either accept or decline a request to help the user with
dementia. The activation of the users starts a \emph{mission}, in which
volunteers and a relative coordinate and work together to bring the user with
dementia to a \emph{safe place}, e.g., a care home, a police station, or the
like. Throughout the mission, the location of the user with dementia is
broadcasted to all other involved users. This is done to enable one or more of
the involved users to be able to navigate to the user with dementia and assist
them. When the user with dementia has been brought to a safe place the mission
is closed by one of the involved users.

The \textit{recreational activity calendar} is at the most basic
level, just a calendar where relevant activities can be posted by
volunteers. The calendar also functions as an
intermediary entity that tries to create opportunities for volunteers
and users with dementia to participate in activities together. This is
done by matching volunteers to activities based on submitted activity
preferences and suggesting relevant activities to users with dementia
based on the keywords that they use when searching through the
activity calendar. Whenever a user with dementia navigates to the
description of an activity that he/she finds interesting, one or more
volunteers will be suggested to the user with dementia as a possible
accompanying person for the activity. It is then up to the user with
dementia to take contact with one of the possible suggested volunteers.

\section{Preliminaries} \label{sec:preliminaries}
Several architectural models exist to design and develop applications, but the
\emph{microservice} and \emph{serverless computing} architectural models are
especially interesting for us as they provide an excellent framework for
building a \emph{distributed system} that is maintainable, efficient and that
can easily scale~\cite{devops_handbook}. In the following, we first give a brief
overview of these architectural models and then describe the concrete
technologies adopted for the development of \tool.

\subsection{Microservices \& Serverless Computing}

The microservice architectural model allows a developer to manage and deploy
several smaller autonomous and modular pieces that communicate through an API
via network messaging protocols. As a result of using this type of architectural
model, it can be easier to horizontally scale software components that make up a
system. These modular pieces usually run in their own isolated environment
possibly using their own technology stack.

The popularity of the microservice architectural model has grown hand in hand
with the increased adoption and usage of application containerization.
Containers are lightweight virtualization runtime environments that allow a
developer to run an application locally for testing purposes, as well as in a
production environment on a server. They present a consistent and portable
software environment, where all dependencies can be packaged as a deployable
unit. Multiple application components that run in containers can be managed by
orchestration tools such as Docker Compose~\cite{docker_compose},
Kubernetes~\cite{kubernetes}, or Docker Swarm~\cite{docker_swarm}.

Serverless computing is a relatively new emerging architectural model that
embraces the concept of Function as a Service (FaaS). The concept of FaaS
encompasses the idea that developers should be able to manage and deploy
independent functions that execute in response to events that are triggered by
internal and/or external sources of a system.
Serverless functions are deployed as containers that are stored and managed in
private or public registries (e.g.,~Docker Hub~\cite{docker_hub}). The execution
of a function means pulling and running one or more corresponding containers
from the registry.


Serverless computing comes with many advantages scalability-wise, but certain
inherent problems are also associated with the type of execution model.
More precisely, retrieving a container image and starting a container can take
some time, i.e. it adds additional overhead to the function call. This
phenomenon is usually referred to as a \emph{cold start} and, to avoid it,
running containers are often reused for subsequent invocations of the same
serverless function.
%


\subsection{Technologies} \label{sec:technologies}


At the core of the backend system, \tool uses the \emph{k3s} Kubernetes
distribution by Rancher Labs~\cite{k3s_rancher} to deploy and run all
microservices. K3s is a lightweight certified Kubernetes
distribution\cite{k8s_conformance_tests} designed for production workloads and
optimized for resource-constrained environments (e.g., small servers or IoT
appliances). The system requirements are less demanding in comparison to those
of a traditional Kubernetes cluster, but the same core functionality can be
expected. In this context, k3s makes it possible to run workloads on the master
node controlling the system as well as having the possibility to easily add
additional worker nodes or even other master nodes to create a \emph{high
availability} cluster. With k3s it is thus possible to easily scale from a
single computing node to several. A contributing factor to the simplicity of k3s
is that, with some exceptions, only the most essential components for running a
bare-minimum high availability Kubernetes cluster have been included into the
distribution of k3s.
Moreover, another noteworthy advantage with k3s is that it is possible to
package k3s into a single binary. This greatly simplifies the process of
installing, configuring, and updating a production Kubernetes cluster. 

Most of the microservices running in the Kubernetes cluster have been developed
using the Django REST framework~\cite{django_rest}. The Django REST framework is
an extension built on top of Django, one of the most popular open-source web
development frameworks written in the Python language. The reason for choosing
this framework is primarily because of its wide adoption and maturity (the
initial release of Django was in 2005). Another reason for choosing this
framework is that a lot of different technologies integrate well with the
framework since Django supports a wide variety of databases, caching solutions,
and programming interfaces.
%
To store the microservices data, PostgreSQL databases are used. PostgreSQL
databases utilize 
a host-based persistent volume and are created automatically using the storage
application of k3s.

For implementing the data processing pipeline, \tool relies on the OpenFaaS
serverless computing framework~\cite{openfaas}. OpenFaaS provides a platform for
easily scaling CPU-bound computations implemented as functions.
OpenFaaS comes with default auto-scaling based on the number of requests per second.
Functions can be invoked asynchronously via message brokers, as well as
synchronously through simple HTTP requests. OpenFaaS also has a large community
and supports a wide range of programming languages with a lot of pre-defined and
useful templates to choose from. 

Finally, \tool also uses Redis~\cite{redis}, i.e., an in-memory distributed data
structure store that can be used as a message queue and broker, but also as a
Key-Value store and cache system.

\section{Implementation} \label{sec:implementation}
\begin{figure}
    \centering
    \includegraphics[width=0.5\textwidth]{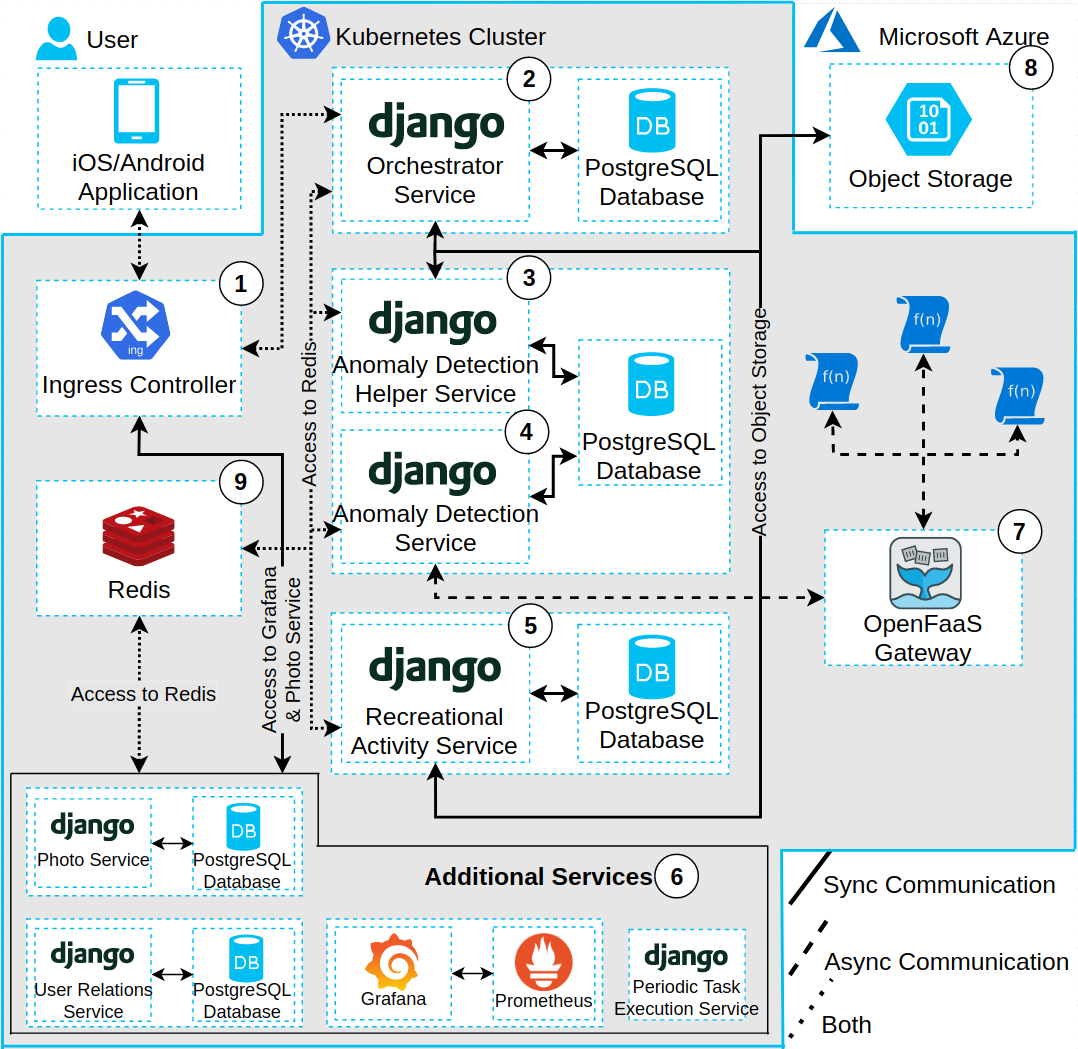}
    \caption{A simplified overview of \tool architecture.}
    \label{fig:architecture}
\end{figure}

Figure~\ref{fig:architecture} gives an overview of the architecture of the
backend of \tool that provides all needed services to the iOS/Android frontend.
Components \circled{2}-\circled{6} are microservices and component~\circled{7}
is the OpenFaaS serverless computing provider.

The entry point of the entire system is an Ingress Controller~\circled{1} that
receives requests from the smartphone application (frontend). Ingress is a
standard component of the Kubernetes platform that provides routing rules to
manage external users' access to the services in a Kubernetes cluster. The
Ingress Controller redirects most traffic to the Orchestrator~\circled{2} but
also reserves a direct route to a Django Photo Service whose sole purpose is to
serve photos such as user profile photos efficiently. Through the Ingress
Controller, a cluster administrator is also able to access monitoring services
such as Prometheus and Grafana.

The Orchestrator microservice~\circled{2} is responsible to route the requests
to other microservices but also to handle the creation, activation, deletion,
update, and retrieval of users. Moreover, it allows to forward notifications to
the smartphone by handling the WebSocket connections, i.e., persistent
connection for bi-directional real-time messaging to the smartphone.

The Orchestrator serves as the logical entry point for all requests to the
\tool's API. It was created to have a unique entry point to
ease the task of monitoring, logging, and caching of requests and to handle user
authorization and authentication in one place only using JSON Web Token
(JWT)\cite{jwt}. This means that a user only needs to authenticate once and then
all internal communication between microservices can happen without further
authentication requests.

Beyond the Orchestrator the backend consists of other main microservices
implementing the functions discussed in Section \ref{sec:sammen_om_demens}. In
particular, the Anomaly Detection Service is implemented by microservices
\circled{3} and \circled{4} with \circled{3} handling bulk operations on raw
location data and acting as a buffer to avoid many sequential writes to the
database, and \circled{4} handling the coordination of OpenFaaS function
execution.  The Recreational Activity Service is implemented by \circled{5}.
Other utility microservices are also used and displayed at the bottom of
Figure~\ref{fig:architecture}.
\begin{itemize}
\item 
The User Relations Service manages relations between users. It handles the
linking between users with dementia and their relatives.  Users' permissions and
prioritization are also handled by this service, which ultimately determines in
which order a relative is contacted in case a related user with dementia has
lost his/her way.  Finally, it handles the invitation sent by one user to
another in the recreational activity component.

\item 
The Photo Service manages images like user profile photos or photos related to
activities posted in the recreational activity calendar. The Photo Service
interfaces with a Microsoft Azure Object Storage~\circled{8} and enables
features for dynamically re-sizing and cropping photos. Furthermore, this
service also works as a webserver for caching and efficiently serving image
data.
\item 
The Periodic Task Execution Service manages periodic tasks such as pushing
database snapshots to the Azure Object Storage, sending out occasional reminders
to volunteer users that have made themselves unavailable, removing users who did
not complete the registration procedure, etc.
\end{itemize}

For logging and monitoring the microservices, \tool relies on the standard
logging capabilities of the Django REST framework.
Prometheus\cite{prometheus} is instead deployed for scraping the usage metrics
of the Kubernetes cluster and its REST API that are visualized for monitoring
purposes using Grafana~\cite{grafana}. The autoscaling of the microservices is
configured based on CPU utilization. Whenever the load is increasing (resp.
decreasing) for a small period of time (10 seconds) then the autoscaling
mechanism is triggered by starting (removing) one or multiple replicas of a
service.

Another central component of \tool's backend is the OpenFaaS
Gateway~\circled{7}. This is the entry point to the OpenFaaS framework that
allows running the functions. OpenFaaS provides the infrastructure for
implementing the algorithms to detect whether a user with dementia has lost the
way. This includes
different artificial intelligence techniques for online and
offline anomaly detection.
The functions deployed to OpenFaaS are developed using Python and are triggered
based on the size of the individual queue in Redis. In Redis, every user has its
own queue when sending location data to the backend. The functions are triggered
asynchronously and can run for an unlimited amount of time. The cold start
problem is avoided by keeping functions warm while there is still a need for
them. The default autoscaling capabilities of OpenFaaS are used to ensure only
the necessary resources are used by considering the number of requests per
second.

The Microsoft Azure Object Storage component \circled{8} is used for storing
photos, binary machine learning model data, database snapshots, and other larger
files. We decided to use Microsoft Azure Object Storage because of its easy
access via the deployment platform Microsoft Azure Cloud, but other equivalent
storing solutions can easily be supported.

Finally, the Redis component~\circled{9} is used as a message queue and broker
in handling the internal communication between microservices.
Two types of communication schemes have been implemented: \emph{synchronous} and
\emph{asynchronous}. The communication scheme mostly depends on the origin of
the initial event.  Graphically, in Figure~\ref{fig:architecture}, bold arrows
indicate synchronous communication and dashed lines indicate asynchronous. The
dotted arrows instead indicate that the services use both synchronous and
asynchronous communications with each other.

Synchronous communication is used whenever an immediate response is required and
must be returned to the requesting entity. This communication scheme is used by
all services that expose some functionality through the Orchestrator and that
can be accessed by a user. For example, when activities in the recreational
activity calendar are created, changed, or retrieved, then synchronous
communication is used. This communication scheme is inherently synchronous as
all tasks will have to be handled within the request-response cycle and if this
is not possible then an error message will have to be returned.

\begin{figure}[t]
    \centering
    \includegraphics[width=0.5\textwidth]{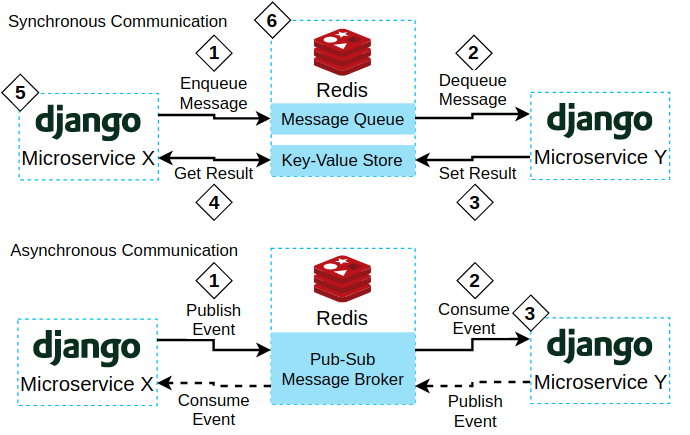}
    \caption{Synchronous and Asynchronous communication schemes between microservices.}
    \label{fig:inter_service_communication}
\end{figure}

The general sequence of actions occurring when transporting synchronously a
message between two microservices is depicted in
Figure~\ref{fig:inter_service_communication} and goes as follows:
\begin{enumerate}
    \item Microservice X enqueues a message with a Universal Unique Identifier
    (UUID) on a named queue.
    \item Microservice Y dequeues and processes the message which in turn
    creates a result.
    \item Microservice Y sets the result in the Redis KV store using the UUID as
    the key and the result as the value.
    \item Microservice X waits for the result to become available under the UUID
    in the Redis KV store.
    \item Microservice X returns or utilizes the result further if available. If
    the result is not available after a number of re-tries and within a certain
    period of time, then the failure of Microservice Y is handled gracefully,
    i.e., a default value and an error message are returned.
    \item The KV pairs stored in the Redis KV store are eventually deleted after
    a small pre-defined Time To Live (TTL) value.
\end{enumerate}

Asynchronous communication is used instead whenever an immediate response is not
required. This communication scheme between microservices is used internally for
functionalities that are not exposed directly through the Orchestrator service.
In general, any interaction that happens outside of the HTTP request-response
cycle is handled asynchronously. The primary place where asynchronous
communication is used is between microservices that communicate with the anomaly
detection service and the service that handles scheduled tasks for future
execution. Instead, any result that needs to be communicated from the backend to
the user is sent via the user's WebSocket connection to the backend or via
a push notification.

The sequence of actions occurring when transporting asynchronously a message
between two microservices is depicted in the bottom half of
Figure~\ref{fig:inter_service_communication} and goes as follows:
\begin{enumerate}
    \item Microservice X publishes an event that needs to be processed by
    Microservice Y.
    \item Microservice Y receives the event as it subscribes to a topic that
    Microservice X publishes on.
    \item Microservice Y consumes and processes the event. Subsequently it
    either does nothing or the result triggers Microservice Y to publish another
    event that Microservice X picks up (indicated by dashed lines in
    Figure~\ref{fig:inter_service_communication}). For example, if the anomaly
    detection service (Microservice Y) detects an anomaly, then an event is
    published for the Orchestrator (Microservice X) that can process it. The
    Orchestrator then sends a WebSocket message or push notification to the
    user. 
\end{enumerate}

Overall these communication schemes can be said to contribute to a loose
coupling of the microservices as they make it easy to add new components and
existing microservices will just have to subscribe to another topic or process
messages from another named queue, to receive events or messages for further
processing.

\section{Performance Evaluation} \label{sec:performance_evaluation}

We set up experiments to test the capabilities of the backend system with
respect to scalability and its ability to handle a number of scenarios with
increasing, decreasing, and varying demand.

\begin{figure}
    \centering
    \includegraphics[width=0.5\textwidth]{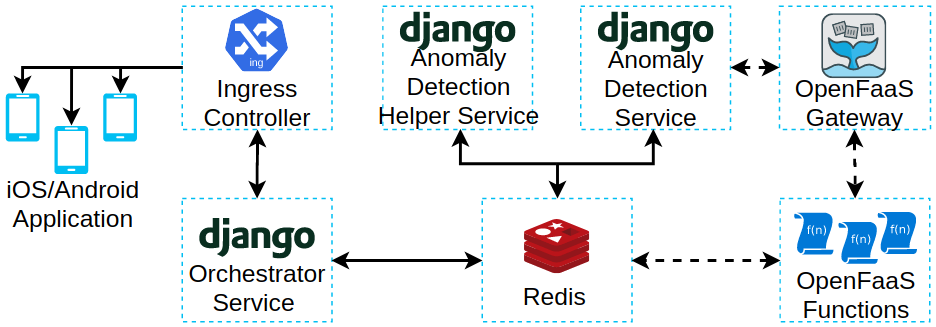}
    \caption{An overview of the anomaly detection part of the backend. Bold arrows indicate synchronous communication and dashed arrows
    indicate asynchronous communication.}
    \label{fig:loadtest_scope}
\end{figure}

The load tests are implemented to target the part of the backend
system that is responsible for providing the anomaly detection
functionality.  Figure~\ref{fig:loadtest_scope} summarizes that part
of the backend system and the communication flows tested.  The
frontend is able to record new locations for a user with a few seconds
granularity (1-5 seconds), whenever the application is running in the
foreground or background of the user's smartphone.
The new locations of the user are sent to the backend system as they
become available through HTTP POST \emph{requests}. These requests
have to be served by the backend within 5 seconds or otherwise a
timeout error is returned. Hence, the backend receives between 12 and
60 new locations per minute. Less demanding operating conditions are
possible. For example, to reduce the impact on battery consumption, it
would be possible to pool locations collected in the frontend at lower
frequency and send the pooled requests to the backend after longer
intervals of time~\cite{RAULT201723}.


In our load tests, we simulate the behavior of real users by means of
virtual users that open the application during a time window of at
most 45 minutes. During this time they send
HTTP POST requests with location data to the backend system every
$\tau$ seconds, with $\tau$ uniformly distributed between 1 and 5. The
data included in a POST request to the backend include a timestamp and
random generated longitude, latitude, altitude, accuracy, speed, and
acceleration.

As shown in Figure~\ref{fig:loadtest_scope}, the location data
eventually arrives at the Anomaly Detection services that aggregate it
and save it in bulk in the PostgreSQL database.  The Anomaly
Detection services also put the location data into a Redis message
queue named after each respective user. After a number of locations
have been placed into the named queue an OpenFaaS function is invoked
and the data in the queue starts being processed by the function until
the size of the named queue falls below a certain limit. The
coordination of the function invocations is handled through Redis as
well.



For the load tests,
the simple task of
computing a moving average of the longitude and latitude coordinates of a user
with respect to a certain window size
is used as a representative example of processing time series data and
take into account time dependencies in the data.
%

More specifically, two different load testing scenarios are designed:
a fixed increasing/decreasing load scenario to test constant and
protracted demand variability and a varying scenario for testing
rapid variability of the number of users.  A steady-state is used as
the initial state in both these two scenarios. It is obtained by adding
users constantly until a load of 2400 virtual users is reached  
and then letting the system stabilize for 5 minutes without further
addition or removal of users.
%

The \textit{Fixed Increasing/Decreasing Load Scenario} proceeds as follows:
\begin{itemize}
    \item 6 users per second are added over a time period of 5 minutes
      resulting in an additional demand of 1800 virtual users. This
      level of demand is maintained for a period of 5 minutes to
      stabilize the system;
    \item next, 14 users per second are added over a time period of 5 minutes reaching a total of
    8400 virtual users. A period of 5
    minutes follows with no change in the level of demand;
    \item next, 14 users per second are removed over a
    period of 5 minutes, such that 4200 virtual users remain. A period of 5 minutes then follows with no change in the level of
    demand; 
    \item finally, 6 users per
    second are removed until only 2400 active virtual users remain. A period of 5 minutes
    follows with no change in the level of demand.  
\end{itemize}

The \textit{Varying Load Scenario} proceeds by varying
more quickly the number of users without waiting for the system to
stabilize. More precisely, it proceeds as follows:
\begin{itemize}
    \item 40 users per second are added over a period of 2.5 minutes.
    This results in a total of 8400 virtual users;
    \item 56 users per second are removed over a period of 2.5
      minutes, which results in 0 virtual users using the backend system.
    \item Immediately after, 56 users per second are added over a period
    of 2.5 minutes. This results in 8400 virtual users using the backend
    system. This step is repeated 6 times. 
    \item next, 40 virtual users per second are removed over a time
      period of 2.5 minutes. This results in a total of 2400 virtual
      users using the backend system concurrently. A time period of 5
      minutes follows with no change in the level of demand.
\end{itemize}

These load tests were written and executed using the Python framework
\emph{Locust}\cite{locust}, i.e., a fairly new load testing framework
and an alternative to more popular but older load testing frameworks
such as Jmeter\cite{jmeter}. Locust was adopted because it is
event-based and it allows a developer to run distributed load tests
with thousands of virtual and concurrent users. It is thus very
suitable for testing highly concurrent workloads, which is also what
the backend system is designed to handle.

The experiments were run on a Kubernetes cluster consisting of one
master node and one worker node deployed on Microsft Azure using
Standard D16v4, Ubuntu 18.04-LTS general-purpose virtual machines with
standard HDD storage. In total, we allocated 32 virtual cores and 64GB
of RAM distributed evenly on two nodes.  All microservices displayed
in Figure~\ref{fig:architecture} were running in different containers
deployed in individual pods.  For each microservice, we configured k3s
autoscaling with a minimum and maximum number of pods along with
resource requests. The configuration was such that at most 20 cores
were used at any time during the experiments by the cluster and 3
cores were used for running the load tests.

\begin{figure}
    \centering
    \includegraphics[width=0.5\textwidth, height=0.4451\textheight]{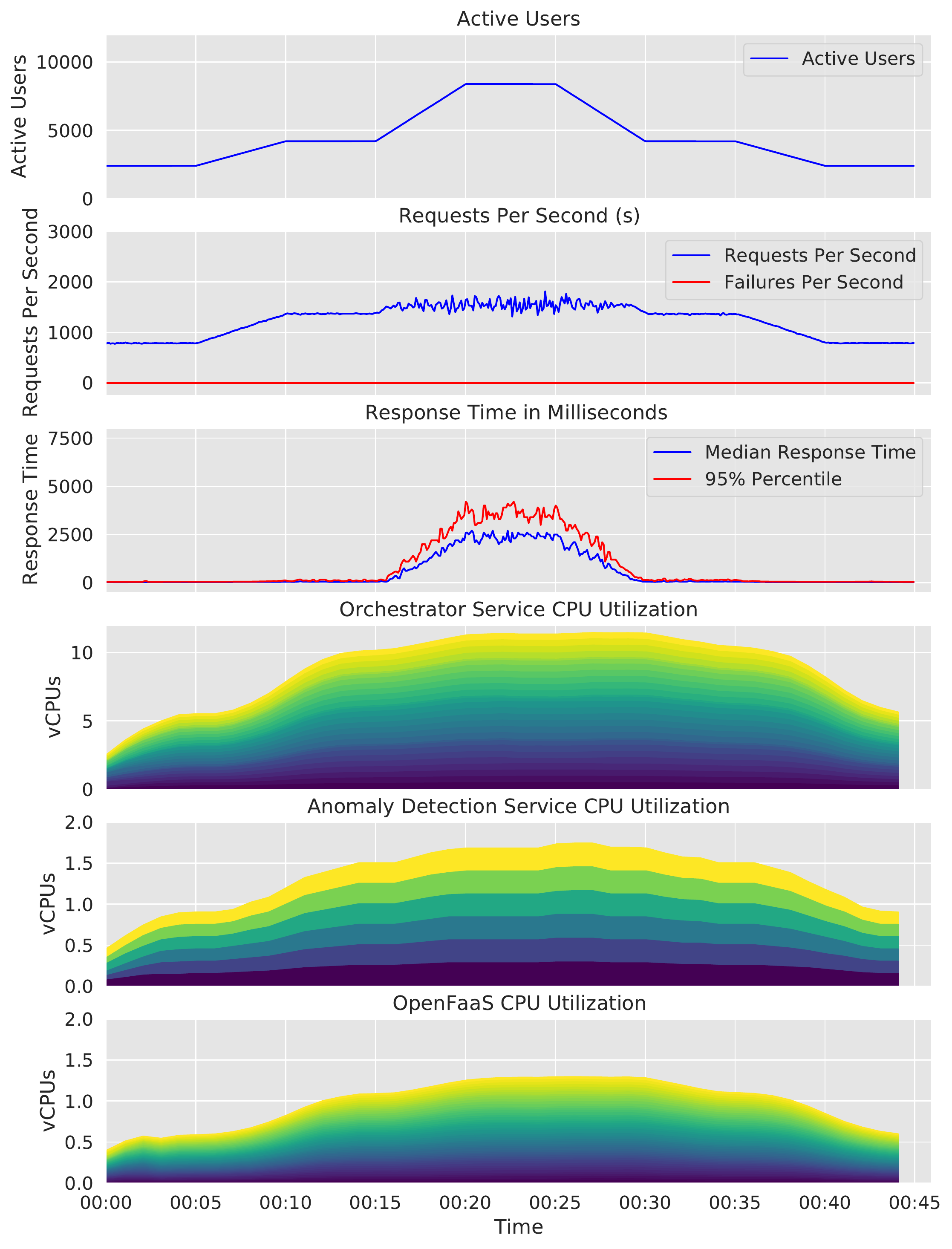}
    \caption{Fixed Increasing/Decreasing Load Scenario.  Each color
      gradient in the area plots correspond to the resources used by a
      replica of a service.}
    \label{fig:loadtest_increasing_decreasing_demand}
\end{figure}

\begin{figure}
    \centering
    \includegraphics[width=0.5\textwidth, height=0.4451\textheight]{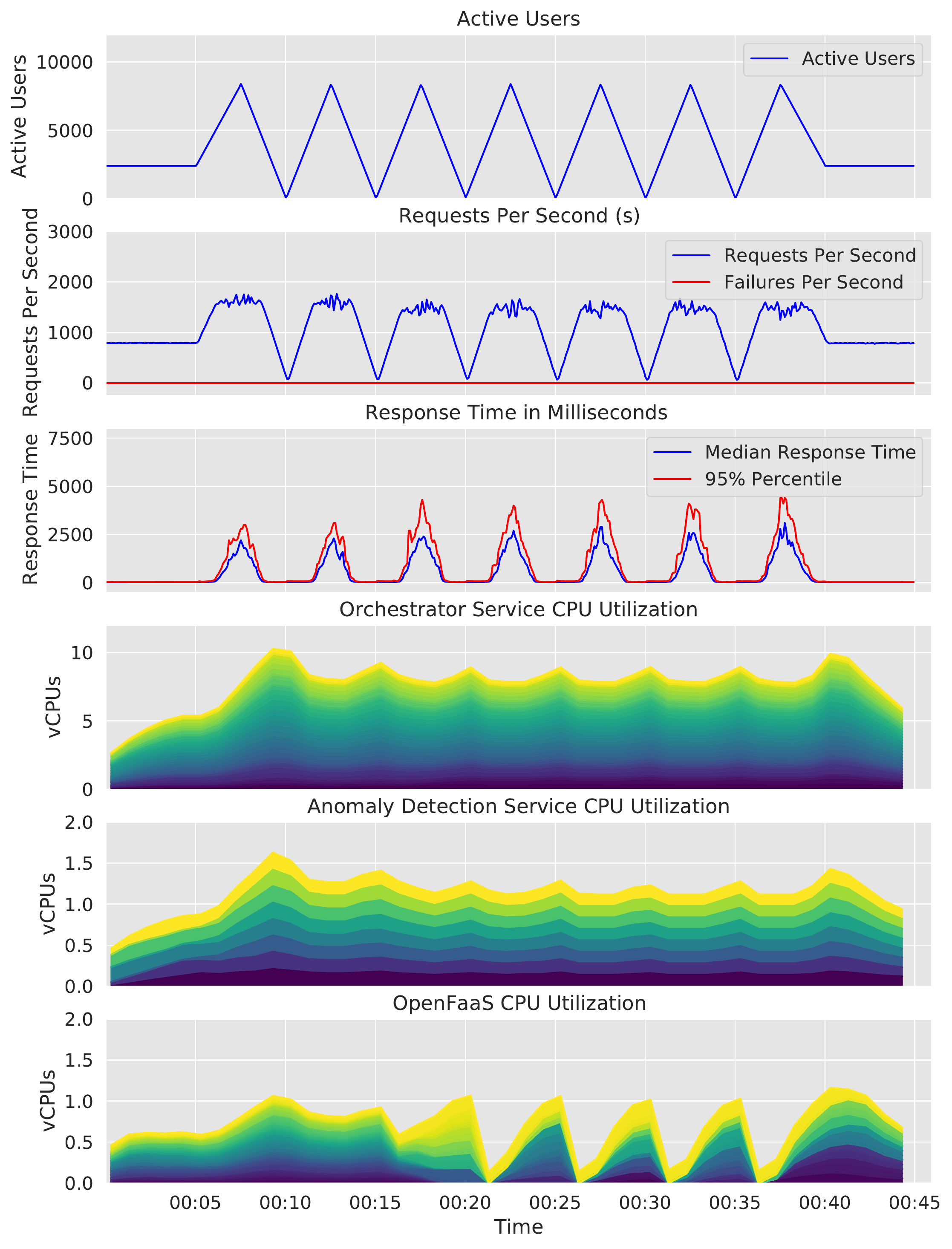}
    \caption{Varying Load Scenario. Each color
    gradient in the area plots correspond to the resources used by a replica of
    a service.}
    \label{fig:loadtest_varying_demand}
\end{figure}

The performance of \tool in the two scenarios is shown in
Figures~\ref{fig:loadtest_increasing_decreasing_demand} and
\ref{fig:loadtest_varying_demand}. Specifically, the plots show the
number of requests, the number of active users, the latency (median
and 95 percentile), and the CPU utilization of certain key components
of the system.  We can see that the services are able to scale
in/out to cope with the different increasing/decreasing and varying
demand patterns. In general, we observe that the response time
increases considerably when the demand peaks. We interpret this as an
indication that the saturation of the resources (i.e., configured max number of pods)
is reached and that the system cannot scale up further. Interestingly,
we never experience failure in responses even in times of high stress
of the system.

The varying load scenario displays the backend system's capability to be
reactive and handle varying demand, while the increasing/decreasing load
scenario displays the backend system's ability to deal with sustained load.

We observe that under the sustained load of the first scenario and the
resources available (the two virtual machines), the number of users
that the backend system is able to handle reliably without an increase in
response time is around 5000 users. Beyond 5000 users we observe that
the response time starts to increase and the number of successfully
served requests per second starts stagnating. A further slow down in
response time beyond 5 seconds will eventually lead to timeouts
because for practical reasons we set the response time limit of the
backend system to 5 seconds. Nonetheless, from the tests in the second
scenario, we learn that the backend system is able to recover from
short periods of time with relatively high demand without any errors.
We observe
that initially during consistent load the
resource usage is distributed evenly among OpenFaaS pods, but during rapid
increases in demand certain pods
are able to start earlier than
others receiving more requests and thus consuming more resources.

In the last three plots of
Figure~\ref{fig:loadtest_increasing_decreasing_demand} and ~\ref{fig:loadtest_varying_demand} we can observe the impact of
the different services on the system. The Orchestrator uses the
largest amount of CPU resources and would be therefore the place to
look at for further optimizations. On the other hand, an insignificant
amount of RAM (not shown in the figures) was used throughout the
tests, which makes sense as it was primarily the throughput of the
system that was tested.

In these tests, we restricted ourselves to two nodes. However, the
backend system can easily be deployed with no further effort on a
larger cluster of nodes thanks to Kubernetes that is designed to scale
to hundreds of computing nodes. Thus, we could handle a much larger
number of active users.

\section{Related Literature \& Conclusions}
\label{sec:related_literature_conclusions}
Smartphone apps, Internet of Things, and wearable sensors offer new
opportunities in the health sector to cure or alleviate diseases as
well as to collect data to gain new knowledge. In line with the
current efforts in the exploitation of these technologies, we designed
a software architecture that allows the collection of data in a
frontend and the analysis in remote in a backend. We focused our
architecture on an app for the detection of getting lost behavior in
people with dementia. However, we believe that the architecture is
reusable in similar data science projects.

We designed the backend system to rely on microservices and serverless
services. Our simulation tests confirmed that the architecture makes it possible to cope
efficiently and reliably with different types of load and to scale appropriately.

Similar to ours, other works on different applications investigate the
performance, scalability, and especially the cost of running microservices and
serverless architectures in the public cloud. They seem to reach similar
conclusions to ours and complement them with additional remarks.
%
%
%
%
%
%
For example, \cite{cho2016} proposes a framework for supporting analytics in
mobile health applications while
%
\cite{mai2013} describes the functionality and architecture of both the frontend
app and the backend system for a smartphone e-health app that supports maternal
health workers in rural India. The work in \cite{bal2016} illustrates instead
the development of three concrete app examples natively on the cloud using
Apache OpenWhisk~\cite{openwhisk}. Like in our work, the authors describe how
easy it is to make changes and deploy functions on the fly.
In \cite{llo2018}, the authors present a case study of the migration of a
monolithic application to AWS Lambda observing great cost savings compared to
running the same application on virtual machines using IaaS.
Similar observations are made in~\cite{ish2018,vil2016}.

Several open-source libraries and frameworks that utilize serverless computing
for running data processing workloads in the cloud have been proposed in the
literature. Most notably, the OSCAR (Open Source Serverless Computing for
Data-Processing Applications) framework is presented in \cite{per2019}. The
framework utilizes an architecture that primarily uses serverless computing for
event-driven data processing. It is a framework for general-purpose
file-processing applications and it uses, like us, OpenFaaS for function
execution. The authors show the capabilities of the developed framework to scale
appropriately according to demand. Furthermore, the reliability and scalability
of the platform is tested on a concrete use-case related to object detection in
videos.

Another framework that aims to simplify the access to distributed computing
through serverless computing is the PyWren framework presented in
\cite{jon2017}. The framework is created to easily perform Python-based
distributed computing on AWS Lambda by avoiding having to provision and
configure complex on-premise clusters. The SIREN machine learning library
proposed in \cite{wan2019} also takes advantage of the same benefits of
deploying to AWS Lambda. The library enables a swarm of stateless functions to
work on batches of machine learning training data. With this approach, the
authors show that they can achieve a high level of scalability.

Several other libraries and frameworks are dedicated to solving problems with
resource provisioning and management in the context of serverless computing. An
example of this is the BARISTA framework introduced in \cite{bhat2019}. It is a
serverless framework that focuses on being able to dynamically manage resources
by horizontally and vertically autoscaling containers based on predicted
workloads. Another framework that is concerned with the same aspect of
serverless computing, but with a more general and broader scope, is presented in
\cite{enes2020}.

\medskip

With the architecture of \tool in place, we are currently working at
the development of the functionalities and the data processing
pipeline in OpenFaaS. In particular, we are deploying several
artificial intelligence techniques to detect if a person with dementia
is getting lost.
%
%
%
The current infrastructure has been extended by coordinating more
function calls and allowing access from the function to other
components of the system, such as the Azure Object Storage. 
%

As far as efficiency is concerned, we noted that compared to other
architectures, our architecture provides a lot of functionality
directly in the Orchestrator Service (e.g., gateway, authorization,
authentication). While handling all these tasks at a single point
might have some advantages, it has of course also the drawback of
requiring more resources to perform them. Hence, looking for possible
further improvements in the scalability and efficiency of the backend
system, we are considering delegating all these functionalities to
other separate services and make them directly accessible through the
Ingress Controller.

Our next steps for the improvement of \tool are: evaluating and combining 
the AI techniques for
detecting getting lost episodes on the basis of benchmark data,
carrying out user tests for assessing the reliability and reception of
the functionalities of the app, and comparing with alternative
proposals from the literature.
Moreover, we plan to complement the location information with other
information coming from sensors on the smartphone and ultimately also
from wearable devices that could remove the need for remembering to
take the smartphone when going for a walk, something we cannot rely on
in advanced stages of dementia.

\section*{Data Availability} \label{sec:data_availability}
A snapshot of the repository containing the codebase for the \tool backend system
and the data that support the findings of this study are openly available in
Zenodo at \cite{data_available}.

\section*{Acknowledgments}

We acknowledge financial support for the project SammenOmDemens by
TrygFonden (ID 127211), Nyborg Kommune and University of Southern
Denmark.

\IEEEtriggeratref{19}
\bibliographystyle{IEEEtran}
\bibliography{IEEEabrv,bibliography}

\end{document}